\documentstyle[11pt,newpasp,twoside,epsf]{article}
\def\edcomment#1{\iffalse\marginpar{\raggedright\sl#1\/}\else\relax\fi}
\reversemarginpar

\newcommand{\lya}{Ly$\alpha$\ }
\newcommand{\halpha}{H$\alpha$\ }
\newcommand{\ha}{H$\alpha$\ }

\begin{document}
\title{The population of galaxies in the forming cluster around the 
radio galaxy MRC 1138-262 at z=2.2}
 \author{L.Pentericci}
\affil{Max-Planck-Institute f\"ur Astronomie, K\"onigstuhl 17,
             D-69117, Heidelberg, Germany}
\author{J.D. Kurk, H.J.A. R\"ottgering, G.K. Miley, B.P. Venemans}
\affil{Sterrewacht Leiden, P.O. Box 9513, 2300 RA, Leiden, 
             The Netherlands}

\begin{abstract}
We have recently discovered  a forming cluster around the radio galaxy MRC 1138-262 at redshift 2.2. Besides the population of \lya\ emitting galaxies  
that have been confirmed spectroscopically, 
we have detected many candidate \ha\ emitters 
that seem to have a different spatial distribution from the other galaxies: they are more clustered towards the center of the cluster and seem to be distributed along the same direction as the radio source.  
We present here the characteristics of the Ly$\alpha$ and H$\alpha$ emitters
and study the nature of these populations.  
\end{abstract}

\section{Introduction}
One of the most interesting questions in modern astrophysics concerns the formation of structures in the early universe. 
The most luminous high redshift radio galaxies are unique probes of galaxy and cluster formation: they are amongst the oldest and most massive objects in the early universe  and there is considerable evidence that they are at the centers of proto clusters.
Here we present results of a search for emission line galaxies in the field around the radio source 1138-262 at z=2.2. Based on the success of this project,
we are now conducting a large program on the VLT to detect and study clusters around a few selected high redshift radio galaxies, over a redshift range from 2 to 4, which is presented by  Kurk et al. in this proceedings.

\section{A forming cluster at redshift 2.2}

We have initially selected the radio galaxy MRC 1138-262 
at z=2.156 as our best field to search for  companion galaxies,
since there are many observations indicating  that it is probably 
a forming massive galaxy at the center of a dense region. 
These include:
its extremely clumpy optical and near-IR morphology, similar to that predicted by simulations of forming massive galaxies (Pentericci et al 1998); 
the extremely distorted radio source, with evidence of interaction 
with the dense surrounding gas; the extreme Faraday rotation of 6600 rad m$^{-2}$, the largest amongst all high redshift radio galaxies known (Carilli et al. 1997), indicating the presence of dense magnetized gas on large scales; and the luminous 
Ly$\alpha$ halo, extending for more than 200 kpc (Kurk et al. 2001).
\\
Initial VLT narrow band imaging centered on the redshifted \lya line
revealed the presence of $\sim$ 50 candidate \lya emitters, 
with restframe equivalent width larger than 20 \AA\ (Kurk et al. 2000). Follow-up 
multi object spectroscopy has  revealed 14 galaxies and 1 faint QSO, at approximately the same distance as the radio galaxy, and at least 4 other objects with spectral energy distribution consistent with them being 
at $z \sim 2.2$ (Pentericci et al. 2000).
All the Ly$\alpha$  emitting galaxies, with the exception of the faint QSO, 
have redshift in the range 2.16$\pm$0.02
and are within a projected physical distance of less than 0.75 h$_{100}^{-1}$ Mpc from the radio source (q$_0 =0.5$). 
There is no  is no clear enhancement of \lya\ emitters toward the radio galaxy,
although the distribution of most of the emitters seems to be elongated in the direction of the radio source (Figure 1).
The velocity distribution suggests that there are two galaxy subgroups, having velocity dispersions of 280 and 500 km s$^{-1}$ respectively and a relative velocity of 1800 km s$^{-1}$ (see Figure 1 in Kurk et al. this proceedings)
\\
Assuming an isothermal sphere model, the mass contained inside a 0.75h$^{-1}$ 
Mpc radius, is respectively 0.3 $\times$ 10$^{14}$h$_{100}^{-1}$ and 0.9 $\times$ 10$^{14}$ h$_{100}^{-1}$ M$_{\odot}$, which implies a total mass in excess of 10$^{14}$ M$_{\odot}$ for the whole system.
\subsection{\ha emitters}
Prompted by the success of our optical narrow band search, we also
obtained near infrared deep images of the 1138-262 field, with ISAAC, 
using a narrow band 
filter centered around 2.07$\mu$m which encompasses \ha emission from galaxies at a redshift around 2.15.  
Since \ha emission is not as sensitive to dust extinction as \lya these observations are mainly sensitive to dusty galaxies.
\\
 Two ISAAC fields
were observed, both inside the FORS field: one centered around the radio galaxy
and one North East of this, covering 6 confirmed Ly$\alpha$ emitters. 
The observations were also complemented by broad band near infrared 
images, in K band for the eastern field and in J-H-K band for the central field
\\
Candidate line emitting objects were then selected on the basis of their excess narrow versus broad band flux.
We have selected objects with restframe equivalent width EW$_0$ $>$ 50 \AA\ and $\Sigma$ $>$ 3, where $\Sigma$ is the excess signal in the narrow band compared
with the noise in both narrow and broad band, as defined in Bunker et al. (1995).
As a result we detect 17 candidates \ha emitters in the central field
and only 6 in the  NE field, with one in common between the two fields (Figure 1).
\begin{figure}
\plotfiddle{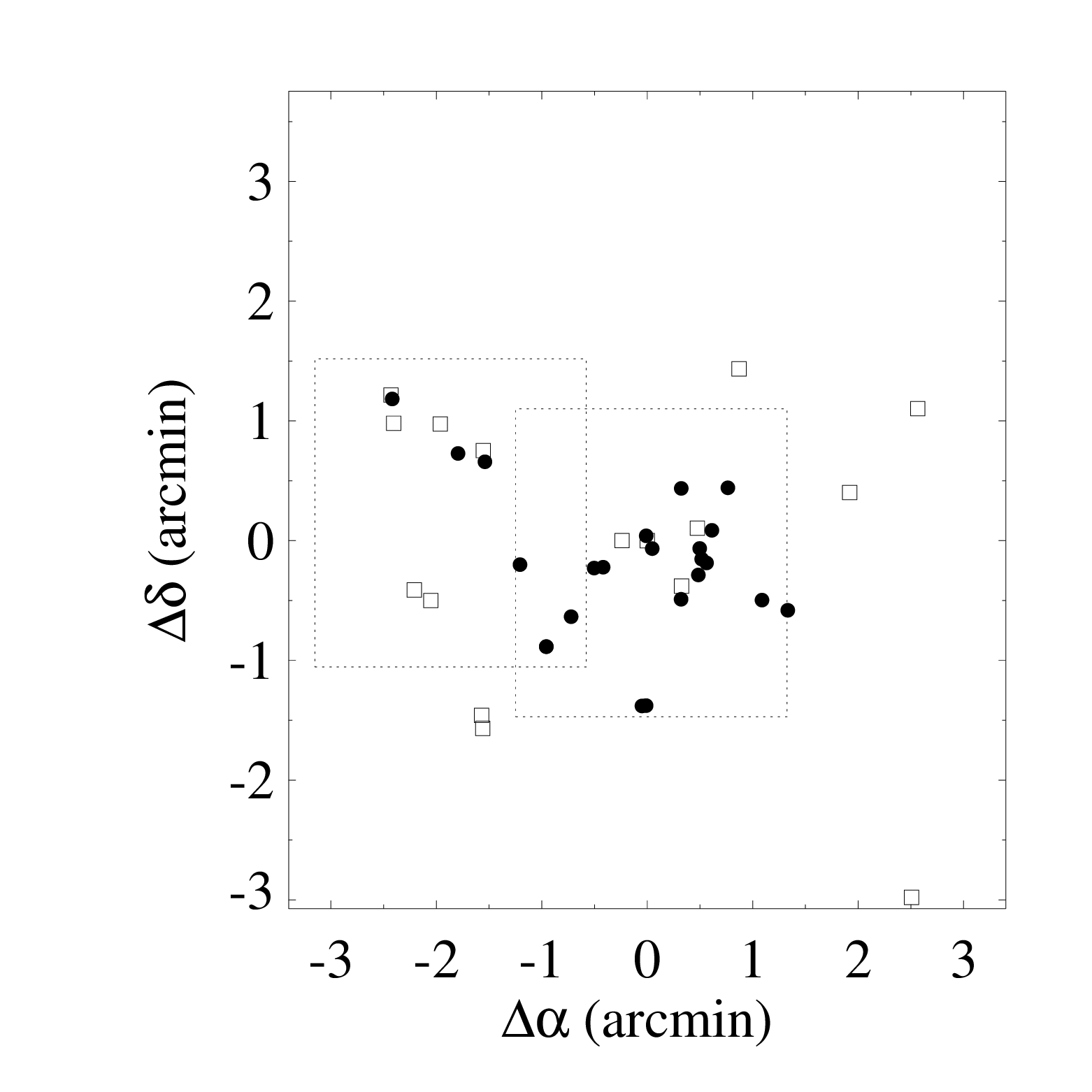}{4cm}{0}{40}{40}{-180}{-230}
\vskip2.4cm
\caption{ Positions of the \lya \ emitters (empty squares) and \ha\ emitters (circles) around the radio galaxy 1138-262. The total field is the FORS1 field of view, while the two dotted areas represent the two ISAAC fields imaged in the infrared.}
\end{figure}
We will assume that the line detected is H$\alpha$. Other possibilities that cannot be excluded, although less likely, include fainter lines from higher redshift objects, such as [OII](3727 \AA) from z=4.6 galaxies or [OIII](5007 \AA) 
from z=3.2 galaxies, and P$\alpha$ (at 1.88$\mu$m) from z=0.12 galaxies. Confirmation of the real redshift is one of the aims of the planned follow-up spectroscopy that will be carried out next March with ISAAC.
\\
The spatial distribution of the H$\alpha$ emitters is clearly not homogeneous over the two fields but rather confined to the central parts where there are 3 times as many candidates as in the outer region.
Moreover in the central field the emitters seem to be 
oriented in a similar direction as the  radio source and the ionized gas halo that surrounds the galaxy. 
This is clearly different from the \lya emitters that show no clustering on this scale.

\section{The galaxy population of the forming cluster}
The population of H$\alpha$ emitters could just be a dusty version of the 
Ly$\alpha$ emitters. Both \lya\ and \halpha\ are produced
 by recombination of ionized
hydrogen and are indicators of star formation rate.  
However, since \lya\ is a resonant line, \lya\
photons will be scattered by neutral hydrogen: if there is also dust present
in such a region, most \lya\ photons will be absorbed before they can escape. 
Therefore, \lya\ emission can
only be observed from galaxies with a low or very patchy dust
content. This is not the case for \halpha\ radiation which 
is less susceptible to dust. Hence, with an
\halpha\ survey we also target  those galaxies 
with high star formation accompanied by larger amounts of dust.
\\
An indication that this is indeed the case comes from 
the ratio of  SFR derived from  \ha emission, and from the UV continuum.
In Figure 2 we present such quantities for all the \ha emitters: 
the SFRs have been derived according to the prescription of Kennicutt (1998), and in particular for the line emission we used SFR(M$_\odot$/yr) = 7.9$\times$10$^{-42}$ L(\ha)(erg s$^{-1}$), which is appropriate for continuous star formation, and for the UV restframe continuum we used
  SFR(M$_\odot$/yr) = 1.4$\times$10$^{-28}$ L(1500-2800\AA)(erg s$^{-1}$ Hz$^{-1}$).
The ratio of the two quantities is around one or larger in almost all cases, with the exception of the brighter 2-3 objects which might as well be lower redshift interlopers.
\begin{figure}
\plotfiddle{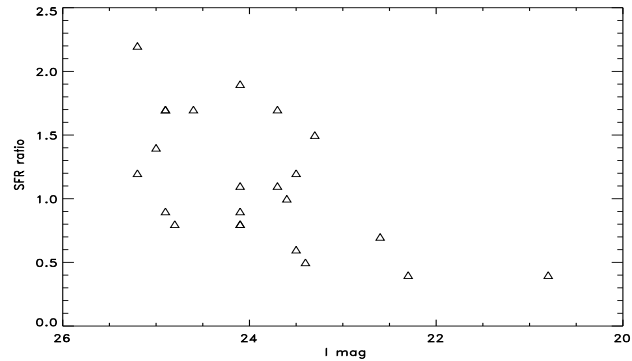}{4cm}{0}{50}{40}{-130}{-10}
\caption{The ratio of SFR derived from \ha emission to the SFR derived from the continuum emission at 2000 \AA\ restframe, versus the total I-band magnitude of the galaxies. For most objects the ratio is well  above 1 indicating the presence of dust.}
\end{figure}
However if the dust content was the only difference between the 
\lya and the \ha emitters, we would not expect a different distribution between the two populations. Possible explanations could then be a difference 
in age and/or metal content.  
We are still investigating this issue: preliminary results 
from the broad band photometry 
seem to indicate no differences between the two populations, both having spectral energy distributions consistent with young starburst galaxies.
\subsection{An enhanced AGN content?}
The mean rest frame EW$_0$ for the Ly$\alpha$ emitters
is 60 \AA\ and the distribution is nearly uniform from 15 to 150 \AA. 
There are several objects with EW in excess of 100 \AA.
This is significantly different from what is obtained by Steidel et al. (2000)
for a large sample of galaxies at redshift around 3: they 
find no galaxies with EW$_0$ larger than 100 \AA. 
The same considerations can be made for the \ha\ emitters which have very high restframe EW$_0$.
While our results are most probably due to our observational
strategy, that emphasizes large  EW$_0$ objects,
 it is also possible that some of our cluster galaxies, besides the faint QSO, 
contain an AGN contribution, 
since photoionization from hot stars is unlikely to
produce such large EW$_0$ as observed (e.g.\ Charlot \&
Fall 1993). 
An enhanced AGN content for the cluster, is also indicated by the
detection of several other faint X-ray emitters in the field 
(from Chandra observations, Carilli et al 2001), of which one coincides with the QSO, and two others with candidate line emitters.

\end{document}